\documentstyle[preprint,aps]{revtex}
\begin{document}
\draft
\newcommand{\beq}{\begin{equation}}
\newcommand{\eeq}{\end{equation}}
\newcommand{\bea}{\begin{eqnarray}}
\newcommand{\eea}{\end{eqnarray}}
\newcommand{\half}{\frac{1}{2}}
\newcommand{\ihalf}{\frac{i}{2}}
\newcommand{\param}{\epsilon}
\newcommand{\opar}{\overline{\eta}}
\newcommand{\ghalf}{\frac{g}{2}}
\newcommand{\quart}{\frac{1}{4}}
\newcommand{\cA}{\cal A}
\newcommand{\cB}{\cal B}
\newcommand{\cH}{\cal H}
\newcommand{\suma}{\sum_{q=1}^2}
\newcommand{\dmu}{\partial_{\mu}}
\newcommand{\dmup}{\partial^{\mu}}
\newcommand{\aslash}{\not\!\!\cA}
\newcommand{\hslash}{\not\!\!\cH}
\newcommand{\Dslash}{\not\!\! D}
\newcommand{\dslash}{\not\!\partial}
\renewcommand{\O}{\Omega}
\renewcommand{\a}{\alpha}
\renewcommand{\t}{\theta}
\renewcommand{\S}{\Sigma}
\newcommand{\g}{\gamma}
\renewcommand{\l}{\lambda}
\newcommand{\luq}{\lambda_1^{(q)}}
\newcommand{\ldq}{\lambda_2^{(q)}}
\newcommand{\pq}{\Phi_{(q)}}
\newcommand{\pqd}{\Phi_{(q)}^{\dag}}
\newcommand{\opq}{\overline{\Psi}_{(q)}}
\newcommand{\psiq}{\Psi_{(q)}}
\newcommand{\duum}{\partial^{\mu}N}
\newcommand{\mps}{\vert\phi\vert^2}
\newcommand{\oSig}{\overline{\Sigma}}
\renewcommand{\ol}{\overline{\l}}
\newcommand{\pau}{\vec{\tau}}
\newcommand{\fiq}{\Phi_{(q)}}
\newcommand{\opsi}{\overline{\psi}}
\newcommand{\oW}{\overline{W}}
\newcommand{\oD}{\overline{D}}
\newcommand{\tsq}{\theta^2}
\newcommand{\otsq}{\overline{\theta}^2}
\newcommand{\ts}{\overline{\theta}{\theta}}
\newcommand{\tu}{\t\sigma^{\mu}\tt}
\renewcommand{\tt}{\overline{\theta}}
\newcommand{\G}{\Gamma}
\preprint{UB-ECM-PF 96/9 ~~~ La Plata-Th 96/07}
\title{Supersymmetric Electroweak Cosmic Strings}
\author{Jos\'e D. Edelstein$^a$ 
\thanks{On leave from Universidad Nacional de La Plata.
e-mail address: edels@ecm.ub.es}  and
Carlos N\'u\~nez$^b$
\thanks{e-mail address: nunez@venus.fisica.unlp.edu.ar}} 
\address{$^a$ Departament d'Estructura i Consituents de 
la Mat\`eria \\
Facultat de F\'{\i}sica, Universitat de Barcelona \\
Diagonal 647, 08028 Barcelona, Spain \\ and \\
$^b$ Departamento de F\'\i sica, Universidad Nacional de La Plata\\
C.C. 67, (1900) La Plata, Argentina}

\maketitle
\setcounter{page}{1}

\begin{abstract}
We study the connection between $N=2$ supersymmetry and 
a topological bound in a two-Higgs-doublet system with an
$SU(2)\times U(1)_Y\times U(1)_{Y^{\prime}}$ gauge group.
We derive the Bogomol'nyi equations from supersymmetry considerations
showing that they hold provided certain conditions
on the coupling constants, which are a consequence of the huge 
symmetry of the theory, are satisfied. Their solutions, which can be 
interpreted as electroweak cosmic strings breaking one half of the
supersymmetries of the theory, are studied. Certain interesting 
limiting cases of our model which have recently been considered
in the literature are finally analyzed.
\end{abstract}
\pacs{11.27.+d, 12.60.Jv, 98.80.Cq} 

\section{Introduction}

Supersymmetric Grand Unified Theories (SUSY GUTs) have
attracted much attention in connection with the hierarchy problem 
in possible unified theories of strong and electroweak
interactions \cite{R,PDB}.
In view of the requirement of electroweak symmetry breaking, 
these models necessitate an enrichment of the 
Higgs sector \cite{NACh}, 
thereby raising many interesting questions both 
from the classical and the quantum point
of view. In particular, many authors 
have explored the existence of stable electroweak vortex 
solutions in a variety of multi-Higgs systems \cite{DSPEJBTT,GH,BL} 
that mimic the bosonic sector of SUSY GUTs, in correspondence 
with what happens in the abelian
Higgs model \cite{NO}. It has also been argued that GUT cosmic
strings may exhibit superconducting properties \cite{Pet}, and
this fact has recently stimulated the study of several multi-Higgs 
models
describing many interesting phenomena \cite{RDM,Morr}.

Vortices emerging as finite energy solutions of gauge 
theories can be usually shown to satisfy a topological bound for 
the energy, the so-called Bogomol'nyi bound \cite{BdVS,dVS}. 
Originally, these bounds were obtained by 
writing the energy of the configuration
(per unit length) as a sum of squares plus a topological term. 
There exists another approach to study the Bogomol'nyi relationships
(i.e. {\it Bogomol'nyi bound and equations})
which exploits the huge symmetry of the theory:
it is based on the observation that Bogomol'nyi bounds reflect the
presence of an extended supersymmetric structure 
\cite{WO,LLW,HS,ENS}.
In particular, for gauge theories with 
spontaneous symmetry breaking and a  topological charge, 
admitting of 
an $N=1$ supersymmetric version, it was shown that  
the N=2 supersymmetric extension,                     
which requires certain conditions on coupling constants, 
has a central charge coinciding with the topological charge 
\cite{HS,ENS}.
Having originated from the supercharge algebra, the bound is expected
to be quantum mechanically exact. 

Since multi-Higgs models can be understood to be motivated by SUSY 
GUTs, Supersymmetry provides 
a natural framework for studying Bogomol'nyi bounds. In fact,
we have recently considered in Ref.\cite{ENS2} the supersymmetric 
extension of the
two-Higgs model first presented in \cite{BL}, showing that 
Bogomol'nyi
equations are a direct consequence of the underlying $N=2$
supersymmetry of the model.
We shall study in this letter a supersymmetric
formulation of an $SU(2)\times U(1)_Y\times U(1)_{Y^{\prime}}$ 
model with two-Higgs doublets which is a generalization of the one
analyzed in \cite{ENS2}. 
The theory has the same gauge group structure as
that of supersymmetric extensions of the 
Weinberg-Salam Model that arise
as low energy limits of $E_6$ based Grand Unified theories
or $E_8 \times E_8$ superstring theories 
compactified on a Calabi-Yau manifold with an 
$SU(3)$ holonomy. This gauge group was recently considered in 
Ref.\cite{GH} for the study of electroweak strings and,
generically, the inclusion of an extra $U(1)$ factor in
multi-Higgs systems has been also taken into account in a
variety of models exhibiting cosmic strings \cite{Pet,RDM}. 
In spite of being a simplified model (in the sense that its Higgs
structure is not so rich as that of Grand Unified theories), 
it can be seen as the simplest extension of the Standard Model 
necessary for having the Bogomol'nyi equations. 
We show that the Bogomol'nyi bound of the model, as well as 
the Bogomol'nyi equations, are direct consequences of the 
requirement of $N=2$ supersymmetry imposed on the theory. 
vWe also show explicitely that, as a necessary condition for
achieving the $N=2$ model, certain relations between coupling 
constants must be satisfied.
These ``critical values'' of the coupling constants have 
physical relevance; e.g. the required relation between 
coupling constants in the Abelian Higgs model corresponds to the 
limit between type-I and type-II superconductivity 
in the relativistic Ginsburg-Landau model \cite{ENS}.
We discuss the solutions of the Bogomol'nyi equations,
and present some interesting limiting cases.

The paper is organized as follows: in Section II, we present the
$SU(2)\times U(1)_Y\times U(1)_{Y^{\prime}}$ two-Higgs model
in $2+1$ dimensions admitting of non-trivial topological 
configurations
and we embed it in an $N=1$ supersymmetric theory. 
We show that the $N=2$ supersymmetric extension can be obtained 
provided some relations between coupling constants, analogous to
the critical relation appearing in the Abelian Higgs model 
\cite{ENS},
hold. In section III,
we construct the $N=2$ supercharges of the theory, and
compute the corresponding supersymmetry algebra. 
After static configurations are
considered, and restricting our calculations to the bosonic sector,
we find that the Bogomol'nyi relationships appear as a direct
algebraic consequence. 
This fact clarifies in our theory the model-independent analysis
established in Ref.\cite{HS}. We further comment on some interesting
features of the classical field solutions saturating the Bogomol'nyi 
bound. These
could be interpreted as electroweak cosmic strings breaking
half of the supersymetries of the theory. 

Our approach being general and systematic, we finally consider in
Section IV some limiting cases describing various models which have
been recently considered in the literature.

\section{The $SU(2)\times U(1)_Y\times U(1)_{Y^{\prime}}$ $N=2$
Supersymmetric model}
We start with an $SU(2)\times U(1)_Y\times U(1)_{Y^{\prime}}$ 
gauge theory, which is described by the action
\bea
S & = & \int d^3x \left[ - 
\quart \vec{W}_{\mu\nu}\cdot\vec{W}^{\mu\nu}
- \quart F_{\mu\nu}F^{\mu\nu} - \quart G_{\mu\nu}G^{\mu\nu} 
+ \half\vert{\cal D}_{\mu}^{(1)}\Phi_{(1)}\vert^2 
+ \half\vert{\cal D}_{\mu}^{(2)}\Phi_{(2)}\vert^2 \right.\nonumber\\ 
& + & \left. \half (\dmu A)^2 + \half (\dmu B)^2 + 
\half (D_{\mu}\vec{W})^2 - {V}(\Phi_{(1)},\Phi_{(2)},A,B,\vec{W}) 
\right], 
\label{acbos}
\eea
where $\Phi_{(1)}$ and $\Phi_{(2)}$ are a couple of 
Higgs doublets under the $SU(2)$ part of the gauge group,
$A$ and $B$ are real scalar fields and $\vec{W} = W^a\tau^a$ is a 
real
scalar in the adjoint representation of $SU(2)$.
The metric is choosen to be $g^{\mu\nu}=(+ - -)$ and the specific
form of the potential will be determined below.
The strength fields can be written in terms of gauge fields as:
\beq
F_{\mu\nu} = \partial_{\mu} A_{\nu} - \partial_{\nu} A_{\mu} 
\;\;\; , \;\;\; G_{\mu\nu} = \partial_{\mu} B_{\nu} 
- \partial_{\nu} B_{\mu} 
\label{fgmunu}
\eeq
and
\beq
W_{\mu\nu}^{a} = \partial_{\mu} W_{\nu}^a 
- \partial_{\nu} W_{\mu}^a + gf_{abc}W_{\mu}^bW_{\nu}^c, 
\label{Wmunu}
\eeq
while the covariant derivative is defined as:
\beq
{\cal D}_{\mu}^{(q)}\fiq = \left( \partial_{\mu} + 
\frac{i}{2}gW_{\mu}^a\tau^a + \frac{i}{2}\alpha_{(q)}A_{\mu} 
+ \frac{i}{2}\beta_{(q)}B_{\mu} \right) \fiq, 
\;\;\;\;\; \mbox{\scriptsize q=1,2}
\label{dercov}
\eeq
where $g$ is the $SU(2)$ coupling constant 
while $\alpha_{(q)}$ and $\beta_{(q)}$ represents the different 
couplings of $\Phi_{(q)}$ with $A_{\mu}$ and $B_{\mu}$.

A minimal $N=1$ supersymmetric extension of this model is given by an
action which in superspace reads:
\bea
{\cal S}_{N=1} & = &  \half \int d^3x d^2\theta
\left[{\overline{\Omega}}_A\Omega_A + {\overline{\Omega}}_B\Omega_B 
+ {\overline{\Omega}}^a_{\vec{W}}{\Omega}^a_{\vec{W}} 
- \overline{{\cal D}{\cal A}}{\cal D}{\cal A} - \overline{{\cal 
D}{\cal B}}{\cal D}{\cal B} \right.\nonumber\\
& - & \left.\overline{{\cal D}{\cal W}}^a{\cal D}{\cal W}^a 
+ \xi_1{\cal A} + \xi_2{\cal B} + \half\sum_{q=1}^2
\left[(\overline{\nabla^{(q)}\Upsilon_{(q)}})^a
({\nabla^{(q)}\Upsilon_{(q)}})^a \right.\right.\nonumber\\
& + & \left.\left.
i\Upsilon_{(q)}^{\dag}\left(\sqrt{2\luq}{\cal A} + 
\sqrt{2\ldq}{\cal B} 
+ \sqrt{2\l_3}{\cal W}^a\tau^a\right)\Upsilon_{(q)}\right] \right],
\label{supaction}
\eea
where
\beq
{\nabla^{(q)}\Upsilon_{(q)}}= \left({\cal D} + \ihalf 
g\Gamma_{\vec{W}} + \ihalf \alpha_{(q)}\Gamma_A 
+ \ihalf \beta_{(q)}\Gamma_B\right)
\Upsilon_{(q)}.
\label{nablaq}
\eeq

This action is built from a couple of complex doublet superfields 
$\Upsilon_{(q)} \equiv (\Phi_{(q)},\Psi_{(q)},F_{(q)})$, three real 
scalar superfields ${\cal A} \equiv (A,\chi_A,a)$, 
${\cal B} \equiv (B,\chi_B,b)$ and ${\cal W} \equiv 
(W^a,\chi_{\vec{W}}^a,w^a)\tau^a$
and three spinor gauge superfields which in the Wess-Zumino gauge 
read
$\Gamma_A \equiv (A_{\mu},\rho_A)$, $\Gamma_B \equiv
(B_{\mu},\rho_B)$ and $\Gamma_{\vec{W}} \equiv 
\Gamma_{\vec{W}}^a\tau^a = (W_{\mu}^a,\lambda^a)\tau^a$.
$\Omega_A$, $\Omega_B$ and ${\Omega}^a_{\vec{W}}$, are the 
corresponding 
superfield strengths. Concerning $\lambda_1^{(q)}$,
$\lambda_2^{(q)}$, $\lambda_3$, $\xi_1$ and $\xi_2$, they are real
constants whose significance will be clear below.
It must be stressed that fermions $\rho_A$, $\rho_B$, $\chi_A$ and 
$\chi_B$
are Majorana, $\chi_{\vec{W}}^a\tau^a$ and 
$\lambda^a\tau^a$ are Majorana spinors in
the adjoint representation of $SU(2)$, while the Higgsino doublets 
$\Psi_{(q)}$ are Dirac. $A$ and $B$ are real scalar fields and 
$W^a\tau^a$ is an hermitian field in the adjoint representation of 
$SU(2)$. $F_{(q)}$, $a$, $b$ and $w^a$ are auxiliary fields 
which will be eliminated in what follows using their equations of 
motion.
Finally, ${\cal D}$ is the usual supercovariant derivative:
\beq 
{\cal D} = \partial_{\overline{\theta}} + 
i\overline{\theta}\gamma^{\mu}\dmu, 
\label{super}
\eeq
with the $\gamma$-matrices being represented by $\gamma^0 = \tau^3$, 
$\gamma^1 = i\tau^1$ and $\gamma^2 = -i\tau^2$.
                                               
Written in components, action (\ref{supaction}) takes the form:
\beq
{\cal S} = S + {\cal S}_{Fer}
\label{actionn1}
\eeq
where
\bea
{\cal S}_{Fer} & = & \half \int d^3x 
\left[\ihalf\sum_{q=1}^{2}\overline{\Psi}_{(q)}\not\!\!{\cal 
D}^{(q)}\Psi_{(q)} + \ihalf\overline{\Lambda}^a(\not\!\! D\Lambda)^a 
+ \ihalf\overline{\chi}_{\vec{W}}^a(\not\!\! 
D\chi_{\vec{W}})^a + \ihalf\overline{\chi}_A\not\!\partial\chi_A 
\right.\nonumber \\
& + & \left. \ihalf\overline{\rho}_A\not\!\partial\rho_A
+ \ihalf\overline{\chi}_B\not\!\partial\chi_B + 
\ihalf\overline{\rho}_B\not\!\partial\rho_B 
- g f^{abc}W^a\overline{\Lambda}^b\chi_{\vec{W}}^c  
+ ig\opq\Lambda^a\tau^a\pq \right.\nonumber\\
& - & \left. \sqrt{8\l_3}\opq\chi_{\vec{W}}^a\tau^a\pq
+ i\alpha_{(q)}\opq\rho_A\pq - \sqrt{8\luq}\opq\chi_A\pq
\right.\nonumber\\
& + & \left.i\beta_{(q)}\opq\rho_B\pq - \sqrt{8\ldq}\opq\chi_B\pq 
\right.\nonumber\\
& - & \left.\sum_{q=1}^{2}\left(\overline{\Psi_{(q)}}(\sqrt{2\luq}A 
+ \sqrt{2\ldq}B + \sqrt{2\l_3}W^a\tau^a)\Psi_{(q)} \right) + h.c.
\right].
\label{acfer}
\eea
The potential in (\ref{actionn1}) reads:
\[ {V}(\Phi_{(1)},\Phi_{(2)},A,B,\vec{W}) =
\left(\suma\sqrt{\luq}\pqd\pq - \frac{\xi_1}{\sqrt{2}}\right)^2 + 
\left(\suma\sqrt{\ldq}\pqd\pq - \frac{\xi_2}{\sqrt{2}}\right)^2 \]
\beq
\;\;\;\;\;\;\;
+ 2\sum_{q=1}^{2}\left|\left(\sqrt{\luq}A + \sqrt{\ldq}B + 
\sqrt{\l_3}\vec{W}\right)\pq\right|^2 
+ \l_3\left(\suma\pqd\tau^a\pq\right)^2 
\label{invpot}
\eeq

The preceding action (\ref{actionn1}) is invariant under the 
following 
set of $N=1$ supersymmetry transformations with parameter $\eta$:
\[ \delta W_{\mu}^a = -i\overline{\eta}\gamma_{\mu}\Lambda^a, \;\;\;
\delta \Lambda^a = -i{}^{\star}{W^{\lambda}}^a
\gamma_{\lambda}\eta, \;\;\;
\delta W^a = \overline{\eta}\chi_{\vec{W}}^a, \;\;\;
\delta A = \overline{\eta}\chi_A, \;\;\;  
\delta B = \overline{\eta}\chi_B, \]   
\[ \delta A_{\mu} = -i\overline{\eta}\gamma_{\mu}\rho_A, \;\;\;
\delta \rho_A = -i{}^{\star}F^{\lambda}
\gamma_{\lambda}\eta, \;\;\;
\delta B_{\mu} = -i\overline{\eta}\gamma_{\mu}\rho_B, \;\;\;
\delta \rho_B = -i{}^{\star}G^{\lambda}
\gamma_{\lambda}\eta, \]
\[ \delta \pq = \overline{\eta}\psiq, \;\;\; 
\delta \chi_{\vec{W}}^a = 
- \left[\sum_{q=1}^{2}\sqrt{2\l_3}\pqd\tau^a\pq 
+ i(\Dslash W)^a\right]\eta,   \]
\beq 
\delta \chi_A = - \left[\sum_{q=1}^{2}\sqrt{2\luq}\pqd\pq - \xi_1 
+ i\dslash A\right]\eta, 
\label{trafo}
\eeq
\[ \delta \chi_B = - \left[\sum_{q=1}^{2}\sqrt{2\ldq}\pqd\pq - \xi_2 
+ i\dslash B\right]\eta, \]
\[ \delta \Psi_{(q)} = \left[-i\gamma^{\mu}{\cal D}_{\mu}^{(q)}\pq 
- \left(\sqrt{8\luq}A + \sqrt{8\ldq}B + 
\sqrt{8\l_3}W^a\tau^a\right)\pq\right]\eta, \]
where ${}^{\star}{W^{\lambda}}^a$, ${}^{\star}F^{\lambda}$ and 
${}^{\star}G^{\lambda}$ are the dual field strengths
\beq
 {}^{\star}{W^{\lambda}}^a = 
\half\epsilon^{\mu\nu\lambda}{W_{\mu\nu}}^a,
\;\;\; {}^{\star}F^{\lambda} = 
\half\epsilon^{\mu\nu\lambda}F_{\mu\nu} 
\;\;\mbox{and}\;\;\; 
{}^{\star}G^{\lambda} = \half\epsilon^{\mu\nu\lambda}G_{\mu\nu}.  
\label{fdual}
\eeq

Now, in order to impose the $N=2$ supersymmetric invariance of the 
theory, we can consider transformations with a complex parameter 
$\eta_c$ (an infinitesimal Dirac spinor),
since this implies the existence of two supersymmetries \cite{LLW}. 
$\rho_A$, $\rho_B$, $\chi_A$ and $\chi_B$ being real spinors,
we combine them into Dirac fermions $\Sigma_A$ and $\Sigma_B$ given 
by:
\beq
\Sigma_A \equiv \chi_A - i\rho_A \;\;\;\;\;\;\;\;
\Sigma_B \equiv \chi_B - i\rho_B.
\label{sigmas}
\eeq
We also construct a Dirac fermion $\Xi^a$
in the adjoint representation of $SU(2)$
from $\Lambda^a$ and $\chi_{\vec{W}}^a$
\beq
\Xi^a \equiv \chi_{\vec{W}}^a - i\Lambda^a.
\label{cascada}
\eeq
Using the fermion field redefinitions 
(\ref{sigmas})-(\ref{cascada}), 
the fermionic contribution to the action in (\ref{acfer}) can be
rearranged into the following form:
\bea
{\cal S}_{Fer} & = & \half \int d^3x 
\left[\ihalf\sum_{q=1}^{2}\overline{\Psi}_{(q)}\not\!\!{\cal 
D}^{(q)}\Psi_{(q)} + \ihalf\overline{\Xi}^a(\not\!\! D\Xi)^a 
+ \ihalf\overline{\Sigma}_A\not\!\partial\Sigma_A 
+ \ihalf\overline{\Sigma}_B\not\!\partial\Sigma_B  
\right.\nonumber \\
& - & \left. ig f^{abc}W^a\overline{\Xi}^b\Xi^c - 
\sum_{q=1}^{2}\left(\overline{\Psi}_{(q)}(\sqrt{2\luq}A 
+ \sqrt{2\ldq}B + \sqrt{2\l_3}W^a\tau^a)\Psi_{(q)} 
\right.\right.\nonumber\\
& + & \left.\left. \frac{g-\sqrt{8\l_3}}{2}\opq\tilde\Xi^a\tau^a\pq  
- \frac{g+\sqrt{8\l_3}}{2}\opq\Xi^a\tau^a\pq
\right.\right.\nonumber\\
& + & \left.\left.
\frac{\alpha_{(q)}-\sqrt{8\luq}}{2}\opq\tilde\Sigma_A\pq
- \frac{\alpha_{(q)}+\sqrt{8\luq}}{2}\opq\Sigma_A\pq  
\right.\right.\nonumber\\
& + & \left.\left. 
\frac{\beta_{(q)}-\sqrt{8\ldq}}{2}\opq\tilde\Sigma_B\pq      
- \frac{\beta_{(q)}+\sqrt{8\ldq}}{2}\opq\Sigma_B\pq  
\right) + h.c. \right].
\label{acferult}
\eea
Here $\tilde\Xi^a$, $\tilde\Sigma_A$ and $\tilde\Sigma_B$ are the 
charge conjugates (the complex conjugates) 
of $\Xi^a$, $\Sigma_A$ and $\Sigma_B$ respectively.

We shall be mainly interested in purely bosonic backgrounds where
all fermion fields vanish. 
Given a functional ${\cal F}$ depending both on bosonic and 
fermionic 
fields,
it will then be convenient to define ${\cal F}\vert$ for 
\beq
{\cal F}\vert \equiv 
{\cal F}\vert_{\Psi_{(q)},\Sigma_A,\Sigma_B,\Xi^a = 0}.
\label{spb}
\eeq
Under condition (\ref{spb}) the only non-vanishing 
supersymmetric transformations (\ref{trafo}) 
are those corresponding to fermionic fields:
\beq
\delta_{\eta}\Xi^a\vert = -\left[{}^{\star}{W^{\lambda}}^a
\gamma_{\lambda} + \sum_{q=1}^{2}\sqrt{2\l_3}\pqd\tau^a\pq 
+ i(\Dslash W)^a\right]\eta, 
\label{susyuna}
\eeq
\beq
\delta_{\eta}\Sigma_A\vert = -\left[{}^{\star}F^{\lambda}
\gamma_{\lambda} + \sum_{q=1}^{2}\sqrt{2\luq}\pqd\pq - \xi_1
+ i\dslash A\right]\eta, 
\label{susydos}
\eeq
\beq
\delta_{\eta}\Sigma_B\vert = -\left[{}^{\star}G^{\lambda}
\gamma_{\lambda} + \sum_{q=1}^{2}\sqrt{2\ldq}\pqd\pq - \xi_2 
+ i\dslash B\right]\eta, 
\label{susytri}
\eeq
\beq
\delta_{\eta}\Psi_{(q)}\vert = \left[-i\gamma^{\mu}{\cal 
D}_{\mu}^{(q)}\pq - \left(\sqrt{8\luq}A + \sqrt{8\ldq}B + 
\sqrt{8\l_3}W^a\tau^a\right)\pq\right]\eta. 
\label{trafodos}
\eeq

Now, the transformations (\ref{susyuna})-(\ref{trafodos}) 
with complex parameter $\eta_c = \eta e^{-i\alpha}$ 
are equivalent to transformations with real parameter
$\eta$ followed by a phase transformation for fermions:  
\[ \{\Xi^a,\Sigma_A,\Sigma_B,\Psi_{(q)}\} \longrightarrow 
e^{i\alpha}\{\Xi^a,\Sigma_A,\Sigma_B,\Psi_{(q)}\}. \]
Then, $N=2$ supersymmetry requires invariance under this fermion
rotation.
One can easily see from (\ref{acferult})
that fermion phase rotation invariance is achieved
if and only if:
\beq
\l_3 = \frac{g^2}{8} \;\;\;\;\;\;\; , \;\;\;\;\;\;\;
\luq = \frac{\alpha_{(q)}^2}{8} \;\;\;\;\;\;\; \mbox{and} 
\;\;\;\;\;\;\;
\ldq = \frac{\beta_{(q)}^2}{8}.
\label{conds}
\eeq
That is, the model is invariant under an extended supersymmetry 
provided 
relations (\ref{conds}) are imposed. This kind of conditions 
appears in 
general when, starting from an $N=1$ supersymmetric gauge model, 
one attemps to impose a second supersymmetry: 
conditions on coupling constants have to be imposed so as to 
accommodate different $N=1$ multiplets into an $N=2$ multiplet.
We note that the same conditions take place in the model studied in 
Refs.\cite{BL,ENS2}. Moreover, once (\ref{conds}) are imposed, 
the Higgs
potential of our model happens to be a simple generalization of that
obtained in \cite{BL} by a different approach. In our case, however,
it has been dictated just by supersymmetry considerations. As can be 
seen in Ref.\cite{ENS}, this discussion is analogous to that 
in the Abelian Higgs model.

Summarizing, we have arrived to the following $N=2$ supersymmetric 
action
associated to the $SU(2)\times U(1)_Y\times U(1)_{Y^{\prime}}$ 
model of our interest:
\bea
{\cal S}_{N=2} & = & \half \int d^3x \left[ - \half 
W_{\mu\nu}^a{W^{\mu\nu}}^a - \half F_{\mu\nu}F^{\mu\nu} 
- \half G_{\mu\nu}G^{\mu\nu} + \vert{\cal 
D}_{\mu}^{(1)}\Phi_{(1)}\vert^2 + \vert{\cal 
D}_{\mu}^{(2)}\Phi_{(2)}\vert^2  \right. \nonumber\\
& + & \left. (\dmu A)^2 + (\dmu B)^2 + (D_{\mu}\vec{W})^2 - 2 
{V}(\Phi_{(1)},\Phi_{(2)},A,B,\vec{W}) 
\right. \nonumber\\
& + & \left. 
i\sum_{q=1}^{2}\overline{\Psi}_{(q)}\not\!\!{\cal 
D}^{(q)}\Psi_{(q)} 
+ i\overline{\Xi}^a(\not\!\! D\Xi)^a 
+ i\overline{\Sigma}_A\not\!\partial\Sigma_A 
+ i\overline{\Sigma}_B\not\!\partial\Sigma_B 
- g f^{abc}W^a(i\overline{\Xi}^b\Xi^c \right.\nonumber\\
& + & \left. h.c.)
- \sum_{q=1}^{2} \left[\overline{\Psi}_{(q)}(\alpha_{(q)}A 
+ \beta_{(q)}B + gW^a\tau^a)\Psi_{(q)} 
- g(\opq\Xi^a\tau^a\pq + h.c.) 
\right.\right. \nonumber\\
& - & \left.\left. \alpha_{(q)}(\opq\Sigma_A\pq + 
h.c.) - \beta_{(q)}(\opq\Sigma_B\pq + h.c.) 
\right]\right]. 
\eea

In the next section, the reasons why the conditions (\ref{conds}),
that ensure $N=2$ supersymmetry, are also needed for the 
Bogomol'nyi bound will be clear in the light of the supercharge 
algebra.

\section{Supercharge algebra and Bogomol'nyi equations}

We shall now analyze the $N=2$ algebra of supercharges for our 
model. 
To construct these charges we follow the Noether method. 
The conserved current associated with $N=2$ supersymmetry is 
given by:
\beq
{\cal J}_{N=2}^{\mu} =
\sum_{\{\Phi\}}\frac{\delta L}{\delta 
\partial_{\mu}\Phi}\delta_{\eta_c}\Phi +
\sum_{\{\Psi\}}\frac{\delta L}{\delta 
\partial_{\mu}\Psi}\delta_{\eta_c}\Psi -
\theta^{\mu}[\eta_c]
\label{jmu}
\eeq
where $\{\Phi\}$ and $\{\Psi\}$ represent the whole set of 
bosonic and 
fermionic fields respectively.
Concerning $\theta^{\mu}[\eta_c]$, it is defined
through
\beq
\delta_{\eta_c}S = \int d^3x \partial_{\mu}\theta^{\mu}[\eta_c].
\label{thetamu}
\eeq
The conserved charge is obtained from the current (\ref{jmu}) as
\beq
{\cal Q}[\eta_c] = \int d^2x {\cal J}_{N=2}^0, 
\label{charge}
\eeq
this giving the following explicit expression
\bea
{\cal Q}[\eta_c] & = & - \ihalf\int d^2x
\left\{{\Sigma}_A^{\dag}\left[{}^{\star}F^{\lambda}
\gamma_{\lambda} + \sum_{q=1}^{2}\frac{\alpha_{(q)}}{2}\pqd\pq 
- \xi_1 + i\dslash A\right]
+ {\Sigma}_B^{\dag}\left[{}^{\star}G^{\lambda}\gamma_{\lambda} 
\right.\right.\nonumber\\
& + & \left.\left. \sum_{q=1}^{2}\frac{\beta_{(q)}}{2}\pqd\pq 
- \xi_2 + i\dslash B\right]
+ {\Xi}^{\dag a}\left[{}^{\star}{W^{\lambda}}^a
\gamma_{\lambda} + \frac{g}{2}\sum_{q=1}^{2}\pqd\tau^a\pq 
+ i(\Dslash W)^a\right] \right.\nonumber\\ 
& + & \left.
\sum_{q=1}^{2}\psiq^{\dag}\left[-i\gamma^{\mu}{\cal 
D}_{\mu}^{(q)}\pq 
- \left(\alpha_{(q)}A + \beta_{(q)}B + gW^a\tau^a\right)\pq\right]
\right\}\eta_c.
\label{Q}
\eea

Since we are interested in connecting the $N=2$
supercharge algebra with the Bogomol'nyi relationships, 
we assume static configurations with $A_0 = B_0 = W_0^a = 0$, 
and we restrict ourselves to a purely bosonic solution of the theory 
after computing the algebra. We obtain, after some calculations
\beq
\{\bar{\cal Q}[\eta_c],{\cal Q}[\eta_c]\}| = 
2\opar_c\gamma_{0}\eta_cP^{0}
+ \opar_c\eta_c Z
\label{qeqe}
\eeq
where
\bea
P^0 & = & E = \half \int d^2x \left[ \half (W_{ij}^a)^2 + 
\half (F_{ij})^2 + \half G^2_{ij} + 
\suma |{\cal D}_i^{(q)}\Phi_{(q)}|^2 + (\partial_iA)^2
\right. \nonumber \\
& + & \left. (\partial_iB)^2 + (D_i\vec{W})^2 +
V(\Phi_{(1)},\Phi_{(2)},A,B,W^a)\right]
\label{po}
\eea
while the central charge is given by:
\bea
Z & = & - \int d^2x \left[ \half\epsilon^{ij}F_{ij}\left(
\suma\frac{\alpha_{(q)}}{2}\Phi_{(q)}^{\dag}\Phi_{(q)} - 
\xi_1\right) 
+ \half\epsilon^{ij}G_{ij}\left(
\suma\frac{\beta_{(q)}}{2}\Phi_{(q)}^{\dag}\Phi_{(q)} 
- \xi_2\right) \right.\nonumber\\
& + & \left. 
\frac{g}{4}\epsilon^{ij}W^a_{ij}
\suma\Phi_{(q)}^{\dag}\tau^a\Phi_{(q)} 
+ i\epsilon^{ij}\suma ({\cal D}^{(q)}_i\Phi_{(q)})({\cal 
D}^{(q)}_j\pq)^* \right].
\label{zeta}
\eea
In eqs.(\ref{po})-(\ref{zeta}) conditions (\ref{conds}) 
have been already imposed.

Finite energy configurations, require the following asymptotic 
conditions on the fields
\beq
{W_{ij}}^{a},F_{ij},G_{ij},\partial_iA,\partial_iB,D_iW^a,{\cal 
D}_i^{(q)}\Phi_{(q)} \longrightarrow 0
\label{asymp1}
\eeq
whereas the Higgs doublets as well as the scalar fields 
must minimize the potential at infinity
\beq
V({\Phi_{(1)}}_{\infty}, {\Phi_{(2)}}_{\infty}, A_\infty,
B_{\infty}, W^a_{\infty}) = 0.
\label{asymp2}
\eeq
This last equation can be shown to give 
the following asymptotic behaviour for the Higgs doublets:
\beq
{\Phi_{(1)}}_{\infty} = \frac{\phi_0}{\sqrt{2}}
\left( \begin{array}{c} 0 \\ \exp{in_{(1)}\varphi} \end{array} 
\right)
\;\;\; , \;\;\;
{\Phi_{(2)}}_{\infty} = \frac{\phi_0}{\sqrt{2}}
\left( \begin{array}{c} \exp{in_{(2)}\varphi} \\ 0 \end{array} 
\right),
\label{asymp3}
\eeq
and, at the same time, the scalar fields must solve
\beq
\left(\alpha_{(q)}A_{\infty} + \beta_{(q)}B_{\infty} + 
gW^a_{\infty}\tau^a\right){\pq}_{\infty} = 0.
\label{asymp4}
\eeq
The last term of eq.(\ref{asymp1}) leads to expressions for $n_{(1)}$
and $n_{(2)}$ given by
\beq
n_{(1)} = - \half
(gW_{\varphi}^3 + \alpha_{(1)}A_{\varphi} +
\beta_{(1)}B_{\varphi}) \;\;\;\mbox{and}\;\;\;
n_{(2)} = \half
(gW_{\varphi}^3 - \alpha_{(2)}A_{\varphi} -
\beta_{(2)}B_{\varphi}), 
\label{enes}
\eeq
such that
\beq
m \equiv n_{(1)} + n_{(2)} = 
- \half(\alpha_{(1)}+\alpha_{(2)})A_{\varphi} 
- \half(\beta_{(1)}+\beta_{(2)})B_{\varphi} 
\label{top}
\eeq
is an integer which is inmediately identified as the topological
charge of the configuration.

Coming back to eq.(\ref{zeta}) for the central charge, 
it can be rewritten in the form
\beq
Z = \half \int \partial_i{\cal V}^i d^2x
\label{zetadiv}
\eeq
where ${\cal V}^i$ is given by
\beq
{\cal V}^i =  \left(\xi_1A_j + \xi_2B_j + 
i\suma \Phi_{(q)}^{\dag}{\cal D}^{(q)}_j\pq \right)\epsilon^{ij}
\label{vi}
\eeq
so that, after using Stokes' theorem (and taking into account the
asymptotic behaviours given in (\ref{asymp1})), we obtain
\beq
Z = \oint (\xi_1A_i + \xi_2B_i)dx^i = - 2\pi\phi_0^2 m
\label{zetaint}
\eeq
that is, the central charge of the $N=2$ algebra equals (modulo
some normalization factors) the 
topological charge of the configuration.
This is one of the main points of our work: once the relation between
the central charge and the topological charge is established, a
Bogomol'nyi bound can be easily obtained from the supersymmetry 
algebra \cite{WO,LLW,HS,ENS}.

This sort of identity between the $N=2$ central charge and 
topological 
charge was first obtained by Witten and Olive \cite{WO} in the
$SO(3)$ Georgi-Glashow model. It was 
also discussed for the self-dual 
Chern-Simons system by Lee, Lee and Weinberg \cite{LLW}. 
Hlousek and Spector \cite{HS} have thoroughly analyzed this 
connection by studying several models where the existence
of an $N=1$ supersymmetry and a topological current implies an $N=2$
supersymmetry with its central charge coinciding with the 
topological 
charge. More recently, this connection was established for the
Abelian Higgs model \cite{ENS} where a condition on the coupling 
constants has also been shown to be necessarily imposed. 
This condition is unavoidable both 
for having $N=2$ supersymmmetry and the Bogomol'nyi equations. 
Also, in the study of 
self-dual Chern-Simons systems, having a topological charge 
(related to the magnetic flux) and an $N=1$ extension,
a condition on the symmetry breaking coupling constant must be 
imposed both to achieve $N=2$ extended
supersymmetry and to obtain the Bogomolnyi equations \cite{LLW}.

Coming back to our model, it is now easy to find the Bogomol'nyi 
bound from the corresponding supersymmetry algebra. 
Indeed, since the brackets given 
by (\ref{qeqe}) can be written as a sum of fermionic bilinears,
\bea
\{\bar{\cal Q}[\eta_c],{\cal Q}[\eta_c]\}| & = & \int d^2x \left[
(\delta_{\eta_c}\Xi^a)^{\dag}(\delta_{\eta_c}\Xi^a) + 
(\delta_{\eta_c}\Sigma_A)^{\dag}(\delta_{\eta_c}\Sigma_A) + 
(\delta_{\eta_c}\Sigma_B)^{\dag}(\delta_{\eta_c}\Sigma_B) 
\right.\nonumber\\
& + & \left. 
\suma (\delta_{\eta_c} \Psi_{(q)})^{\dag}(\delta_{\eta_c} 
\Psi_{(q)})\right],
\label{qeqefin}
\eea
it is immediate that
\beq
\{\bar{\cal Q}[\eta_c],{\cal Q}[\eta_c]\}| \geq 0.
\label{cota}
\eeq
This lower bound is saturated if and only if
\beq
\delta_{\eta_c}\Xi^a = \delta_{\eta_c}\Sigma_A = 
\delta_{\eta_c}\Sigma_B = \delta_{\eta_c} \Psi_{(q)} = 0 
\label{cadorna}
\eeq
In order to further analyze the solutions of eqs.(\ref{cadorna}),
let us write the parameter $\eta_c$ as
\beq
\eta_c \equiv \left( \begin{array}{c} \eta_+ \\ \eta_- \end{array} 
\right).
\label{quiral}
\eeq
It is now easy to see that to obtain non-trivial solutions to 
eqs.(\ref{cadorna}) we are forced to choose a parameter 
with definite chirality. Moreover, one can see that the conditions
\beq 
\delta_{\eta_+}\Xi^a = 
\delta_{\eta_+}\Sigma_A = 
\delta_{\eta_+}\Sigma_B = 
\delta_{\eta_+}\Psi_{(q)} = 0
\label{etamas}
\eeq
imply $\delta_{\eta_-}\Xi^a \neq 0$, 
$\delta_{\eta_-}\Sigma_A \neq 0$, 
$\delta_{\eta_-}\Sigma_B \neq 0$, 
$\delta_{\eta_-}\Psi_{(q)} \neq 0$ for nontrivial solutions.
Hence if one is to look for Bogomol'nyi equations corresponding to
non-trivial configurations, it 
makes sense to consider that $\eta_c$ has just one independent
chiral component, say
\beq
\eta_c \equiv \left( \begin{array}{c} \eta_+ \\ 0 \end{array} 
\right).
\label{newquiral}
\eeq
Let us note, at this point, that for a parameter of this form, the
supercharge algebra can be seen to be
\beq
\{\bar{\cal Q}[\eta_+],{\cal Q}[\eta_+]\}| = 
\eta_+^{\dag}\eta_+(2P^{0} + Z)
\label{qeqedos}
\eeq
with $Z$ the central charge whose explicit value is given in 
eq.(\ref{zetaint}).
Then, the inequality (\ref{cota}) is nothing but the Bogomol'nyi 
bound 
of our model
\beq
M \geq \pi\phi_0^2 m.
\label{bogobound}
\eeq
Consequently, eqs.(\ref{cadorna}) are the
Bogomol'nyi equations of the theory (once we identify the 
supersymmetry
parameter with $\eta_+$). Explicitly:
\beq 
\epsilon^{ij}{W_{ij}}^a + g\sum_{q=1}^{2}\pqd\tau^a\pq  = 0 
\;\;\;\;\;\;\; ,
\;\;\;\;\;\;\; (D_{i}W - i\epsilon_{ij}D_{j}W)^a = 0,
\label{bogo1}
\eeq
\beq 
\half\epsilon^{ij}F_{ij} + 
\sum_{q=1}^{2}\frac{\alpha_{(q)}}{2}\pqd\pq - \xi_1 = 0 
\;\;\;\;\;\;\; ,
\;\;\;\;\;\;\; (\partial_{i} - i\epsilon_{ij}\partial_{j})A = 0,
\label{bogo2}
\eeq
\beq
\half\epsilon^{ij}G_{ij}
+ \sum_{q=1}^{2}\frac{\beta_{(q)}}{2}\pqd\pq - \xi_2 = 0 
\;\;\;\;\;\;\; ,
\;\;\;\;\;\;\; (\partial_{i} - i\epsilon_{ij}\partial_{j})B = 0,
\label{bogo3}
\eeq
\beq
({\cal D}_{i}^{(q)} - i\epsilon_{ij}{\cal D}_{j}^{(q)})\pq = 0 
\;\;\;\;\;\;\; , \;\;\;\;\;\;\; 
\left(\alpha_{(q)}A + \beta_{(q)}B + gW^a\tau^a\right)\pq = 0.
\label{bogo4}
\eeq
Owing to (\ref{bogobound}), their solutions also solve 
the static Euler-Lagrange equations of motion. 

Let us remark on the fact that field configurations solving the
Bogomol'nyi equations break half of the supersymmetries, 
a feature common to all models presenting
Bogomol'nyi bounds with supersymmetric extension (See for example
\cite{ENS2} and References therein). 
Indeed, as was seen above, supersymmetry
transformations generated by the antichiral
parameter $\eta_-$ are broken.
If we attempt to keep all the supersymmetries of our model,
we will find that the resulting field configuration has zero energy
(the trivial vacuum) as easily seen from eqs.(\ref{cadorna}).
Had we been faced with an antichiral parameter in (\ref{newquiral}), 
we would have obtained antisoliton solutions with a breaking of the
supersymmetry transformation generated by $\eta_+$.
Analogous results also hold in $4$ dimensional models as the one 
originally studied by Witten and Olive \cite{WO}.
                      
A careful analysis of the whole set of Bogomol'nyi equations, shows 
that it is possible to decouple an equation 
involving only the Higgs doublet in the same vein as it was 
previously done in the abelian Higgs model
\cite{dVS}. 

\section{Limiting Cases}
As a by-product of our systematic approach, we can easily obtain 
Bogomol'nyi bounds (coming from an underlying $N=2$ supersymmetric
structure) for a variety of models which have recently acquired 
physical interest.

\subsection{The $SU(2)\times U(1)_Y\times U(1)_{Y^{\prime}}$ 
pure-Higgs model}
The dynamics of this model first considered in Ref.\cite{BL}, is
dictated by the following Lagrangian density:
\bea
{\cal L}_{pH} & = & - \quart \vec{W}_{\mu\nu}\cdot\vec{W}^{\mu\nu}
- \quart F_{\mu\nu}F^{\mu\nu} - \quart G_{\mu\nu}G^{\mu\nu} 
+ \half\suma\vert{\cal D}_{\mu}^{(q)}\Phi_{(q)}\vert^2 - 
\l_3\left(\suma\pqd\tau^a\pq\right)^2 \nonumber\\ 
& - & \left(\suma\sqrt{\luq}\pqd\pq - 
\frac{\xi_1}{\sqrt{2}}\right)^2 - 
\left(\suma\sqrt{\ldq}\pqd\pq - \frac{\xi_2}{\sqrt{2}}\right)^2. 
\label{acbospH}
\eea
It is immediately seen that the results given in Ref.\cite{BL}
can be obtained just by considering bosonic configurations satisfying
the constraint:
\beq
A = B = W^a = 0. 
\label{set}
\eeq
in our equations\footnote{Note that our algebraic approach is not 
modified by any constraint imposed on purely bosonic configurations,
as all the fermion fields are put to zero 
after computing the algebra.}. 
These conditions are consistent with the asymptotic behaviours
(\ref{asymp1})-(\ref{asymp2}) and with the Bogomol'nyi equations
(\ref{bogo1})-(\ref{bogo4}) of our model.
It is interesting to note
that conditions (\ref{conds}), imposed by the requirement of extended
supersymmetry, also fix in this case
the coupling constants exactly as they appear in
the above mentioned Reference. Thus, we have shown that the 
potential and the coupling constants of the
$SU(2)\times U(1)_Y\times U(1)_{Y^{\prime}}$ pure Higgs 
model studied 
in \cite{BL}, are simply dictated by $N=2$ supersymmetry.
A simple ansatz for string-like solutions of arbitrary topological
charge in this system has been explored in \cite{BL}. It is shown
there that, interestingly enough,  
these configurations do not correspond to an embedding of
the Nielsen-Olesen vortex solution.

\subsection{The $U(1) \times U(1)$ model}
It is well-known that superconducting cosmic strings could 
have appeared as topological defects
in the early Universe, owing to the presence of a charged field 
condensate in the core of the string \cite{W2}. Considering that
supersymmetry could also have been realized in the early Universe,
the study of supersymmetric models possibly involving 
superconducting 
cosmic strings seems to be relevant. As the superconducting
string models are commonly based on a $U(1) \times U(1)$
gauge symmetry \cite{Morr,W2}, 
we will investigate the following Lagrangian density: 
\beq
{\cal L}_{SCS} = - \quart F_{\mu\nu}F^{\mu\nu} - \quart 
G_{\mu\nu}G^{\mu\nu} + \half (D_{\mu}^{(A)}\phi)^*(D^{\mu (A)}\phi)
+ \half (D_{\mu}^{(B)}\xi)^*(D^{\mu (B)}\xi) - V(\phi,\xi)
\label{lagscs}
\eeq
- which can be obtained as a limiting case of our 
$SU(2)\times U(1)_Y\times U(1)_{Y^{\prime}}$ system -,
where $\chi$ and $\xi$ are abelian (complex) Higgs fields, while
$D_{\mu}^{(A)}$ and $D_{\mu}^{(B)}$ are the covariant derivatives
with respect to $A_{\mu}$ and $B_{\mu}$ respectively. To this end, 
we will restrict ourselves to those solutions 
of our model satisfying condition (\ref{set}), which
are decoupled from the non-abelian gauge field $W_{\mu}^a$. That is,
we are interested in topological classical field configurations
in the region of the parameter space where $g \to 0$. We then
ask for solutions where the disconnected non-Abelian field strength
is constrained to vanish,
\beq
W^a_{\mu\nu} = 0.
\label{set3}
\eeq
Now, we rename the abelian coupling constants $\alpha_{(1)}$ and
$\beta_{(2)}$,
\beq
\alpha_{(1)} = \beta_{(2)} \equiv e,
\label{set4}
\eeq
and consider the case where the remaining $\alpha_{(2)}$ and 
$\beta_{(1)}$ vanish.
Finally, we make the following ansatz for the Higgs sector:
\beq
\Phi_{(1)} = \left( \begin{array}{c} \phi \\ 0 \end{array} 
\right) \;\;\;\;\; \mbox{and} \;\;\;\;\;
\Phi_{(2)} = \left( \begin{array}{c} 0 \\ \xi \end{array} 
\right).
\label{set5}
\eeq

After all these steps, our effective Lagrangian looks exactly as 
(\ref{lagscs}), and the explicit form of the Higgs potential reads:
\beq
V(\phi,\xi) = \frac{e^2}{8}(|\phi|^2 - v_1^2)^2 + 
\frac{e^2}{8}(|\xi|^2 - v_2^2)^2.
\label{potscs}
\eeq
Unfortunately, the region of the parameter space spanned by our
Higgs coupling constants, lies outside the range where several 
interesting phenomena take place (See, for example, 
Refs.\cite{RDM,Morr}). To study these, we
would need to modify our starting-point potential.
Nevertheless, let us mention an interesting result that can be
extracted in the model described above. If we consider the 
possibility
that both $U(1)$ symmetries be broken roughly at the same scale,
finite energy leads to the following asymptotic behaviour for the
Higgs fields:
\beq
\phi \to v_1 e^{in_Y\theta} \;\;\;\;\; \mbox{and} \;\;\;\;\;
\xi \to v_2 e^{in_{Y^{\prime}}\theta},
\label{casi}
\eeq  
where $n_Y$ and $n_{Y^{\prime}}$ are integers that characterize the
topological sector to which the solution belongs. Then, in view of
eqs.(\ref{zetadiv}) and (\ref{vi}), it is immediately clear that 
the central charge of the corresponding $N=2$ supersymmetric 
theory becomes:
\beq
\tilde{Z} = - 2\pi v_1^2 \left( n_Y + 
\frac{v_2^2}{v_1^2}n_{Y^{\prime}} \right).
\label{zu1}
\eeq
Thus, the Bogomol'nyi bound of the $U(1) \times U(1)$ model presented
above is
\beq
M \geq \pi v_1^2 \left( n_Y + 
\frac{v_2^2}{v_1^2}n_{Y^{\prime}} \right).
\label{zu2}
\eeq
Remarkably, the bound is not directly proportional to the
topological charge $m = n_Y + n_{Y^{\prime}}$. It would be 
interesting
to investigate how this behaviour matches onto the 
model-independent approach carried out in Ref.\cite{HS}.
 
\subsection{The $SU(2)_{global}\times U(1)_{local}$ semi-local model}
Finally, it is also interesting to explore how $N=2$ supersymmetry 
guarantees the existence of a Bogomol'nyi bound for the neutral
semi-local string defects with $SU(2)_{global}\times U(1)_{local}$ 
symmetry discussed in Ref.\cite{VA}, even though the vacuum 
manifold is 
simply connected. The Lagrangian density of this model takes the
following form:
\beq
{\cal L}_{SL} = - \quart F_{\mu\nu}F^{\mu\nu} 
+ \half [(\dmu - ieA_{\mu})\Phi]^{\dagger}
[(\partial^{\mu} - ieA^{\mu})\Phi] - \l (\Phi^{\dagger}\Phi - v^2)^2,
\label{lagsl}
\eeq
where $\Phi$ is a Higgs doublet charged only under the abelian
subgroup $U(1)_{local}$. The potential is minimum when 
$\Phi^{\dagger}\Phi = v^2$. Since $\Phi$ is a complex doublet,
the minimum of the potential is a three-sphere and is simply 
connected.
This is in contrast with the situation in the abelian Higgs model
where the potential minimum is a circle and a vortex solution
correpond to a configuration which winds around the circle.
However, it was explicitely shown in Ref.\cite{VA} that this
model admits of stable string solutions by a simple generalization
of Bogomol'nyi's proof. We can reproduce their proof as a particular
case of our model. In fact, it is easy to see that imposing 
conditions (\ref{set}) and (\ref{set3}), and working in the
parameter space region where $g,\beta_{(q)} \to 0$, we just have to
restrict ourselves to those configurations satisfying the
following constraint:
\beq
\Phi_{(2)} = G_{\mu\nu} = 0.
\label{sett}
\eeq
Then, the Bogomol'nyi bound obtained in \cite{VA} can be easily
reproduced following the same steps as above.
\vspace{2mm}

Let us end our paper remarking that we have considered 
an $SU(2)\times U(1)_Y\times U(1)_{Y^{\prime}}$ gauge
model with a symmetry breaking potential, which can be seen to be 
a simple extension of the Electroweak Standard model. 
The requirement of $N=2$ supersymmetry
forces a relation between coupling constants and at the same time, 
through its supercharge algebra, imposes the Bogomol'nyi equations 
on certain classical field configurations. 
The connection of our model with realistic supersymmetric extensions 
of the Standard model, and the possible existence of string-like 
solutions in its coupling to supergravity, remain open 
problems. We hope to report on these issues in a forthcoming work.

\section*{Acknowledgements}
This work was partially supported by CONICET, Argentina. We would
like to thank F. Schaposnik for a critical reading of the manuscript.
J.D.E. is pleased to thank the Departament d'Estructura i 
Constituents
de la Mat\`eria for its kind hospitality.

\end{document}